\documentstyle[amsfonts,preprint,eqsecnum,prd,aps]{revtex}
\tightenlines
\begin{document}
\draft
\title{Energy-momentum and angular momentum densities
in gauge theories of gravity}
\author{Toshiharu Kawai\thanks{
Electronic address : kawai@sci.osaka-cu.ac.jp}}
\address{Department of Physics, Osaka City University, 3-3-138 
Sugimoto, Sumiyoshi-ku, Osaka 558-8585, Japan}
\maketitle
\begin{abstract}
In the $\overline{\mbox{\rm Poincar\'{e}}}$ gauge theory of gravity, 
which has been formulated on the basis of a principal fiber bundle 
over the space-time manifold having the covering group of the 
proper orthochronous Poincar\'{e} group as the structure group, 
we examine the tensorial properties of the dynamical energy-momentum 
density ${}^{G}{\mathbf T}_{k}{}^{\mu }$ and the \lq \lq spin" 
angular momentum density ${}^{G}{\mathbf S}_{kl}{}^{\mu }$ of the 
gravitational field. They are both space-time vector densities, 
and transform as tensors under {\em global} $SL(2,C)$-
transformations. Under {\em local} internal translation, 
${}^{G}{\mathbf T}_{k}{}^{\mu }$ is invariant, while 
${}^{G}{\mathbf S}_{kl}{}^{\mu }$ transforms inhomogeneously. 
The dynamical energy-momentum density ${}^{M}{\mathbf T}_{k}{}^{\mu }$ 
and the \lq \lq spin" angular momentum density 
${}^{M}{\mathbf S}_{kl}{}^{\mu }$ of the matter field are also 
examined, and they are known to be space-time vector densities
and to obey tensorial transformation rules under internal 
$\overline{\mbox{\rm Poincar\'{e}}}$ gauge transformations. 
The corresponding discussions in extended new general relativity which 
is obtained as a teleparallel limit of 
$\overline{\mbox{\rm Poincar\'{e}}}$ gauge theory are also given, 
and energy-momentum and \lq \lq spin" angular momentum densities 
are known to be well behaved. Namely, they are all space-time 
vector densities, etc. In both theories, integrations of these 
densities on a space-like surface give the total energy-momentum 
and {\em total} (={\em spin}+{\em orbital}) angular momentum for 
asymptotically flat space-time. The tensorial properties of 
canonical energy-momentum and \lq \lq extended orbital angular 
momentum" densities are also examined.
\end{abstract}
\pacs{04.50.+h}

\narrowtext
\section{INTRODUCTION}
The energy-momentum and angular momentum play central roles in modern 
theoretical physics. The conservation of these is related to the 
homogeneity and isotropy of space-time, respectively. Also, local 
objects such as energy-momentum and angular momentum densities  
are well defined if the gravitational field does not take part in. 
 
In general relativity, however, the energy-momentum and angular 
momentum densities of the gravitational field so far proposed are 
not space-time tensor densities. Rather, it is usually 
asserted \cite{Misner-Thorne-Wheeler} that well-behaved 
energy-momentum and angular momentum densities cannot be defined 
for the gravitational field, while total energy-momentum and total 
angular momentum are defined well for asymptotically flat space-time.

In the $\overline{\mbox{\rm Poincar\'{e}}}$ gauge theory of gravity 
(\=P.G.T.) \cite{Kawai01}, which has been formulated on the basis 
of principal fiber bundle over the space-time manifold having 
the covering group of the proper orthochronous Poincar\'{e} group
as the structure group, we have defined dynamical energy-momentum 
and \lq \lq spin" angular momentum densities. For the asymptotically 
flat space-time, the integration of the dynamical energy-momentum 
density over space-like surface $\sigma $ is the generator of 
{\em internal} translation and gives the total energy-momentum of 
the system. Also, the integration of \lq \lq spin" angular momentum 
density over $\sigma $ is the generator of {\em internal} $SL(2,C)$-transformations and gives the {\em total} (={\em spin}+{\em orbital}) 
angular momentum\cite{spin}, when the Higgs-type field $\psi^{k}$ is 
chosen to be $\psi^{k}=e^{(0)k}{}_{\mu }x^{\mu}+
\psi^{(0)k}+O(1/r^{\beta})$ with constants 
$e^{(0)k}{}_{\mu }, \psi^{(0)k}$
\cite{Kawai02,Kawai-Saitoh01,Kawai-Saitoh02}. In extended new general 
relativity (E.N.G.R.) which is obtained as a teleparallel limit of 
\=P.G.T., corresponding results have been obtained\cite{Kawai-Toma}.

The purpose of this paper is to examine, both in \=P.G.T. and in 
E.N.G.R., the transformation properties of the dynamical 
energy-momentum densities, \lq \lq spin" angular momentum densities, 
canonical energy-momentum densities, and \lq \lq extended orbital 
angular momentum" densities under general coordinate transformations 
and under $\overline{\mbox{\rm Poincar\'{e}}}$ gauge transformations
\cite{Poincare}. The main result is that {\em all the dynamical 
energy-momentum densities and \lq \lq spin" angular momentum densities 
in these theories are true space-time vector densities.}
\section{$\overline{\mbox{\rm POINCAR\'{E}}}$ GAUGE THEORY}
\label{sec:P.G.T.}
\subsection{Outline of the theory}
\label{subsec:outline-PGT}
\=P.G.T. is formulated on the basis of the principal fiber bundle 
${\cal P}$ over the space-time $M$ possessing the covering group 
$\bar{P}_{0}$ of the proper orthochronous Poincar\'{e} group as the 
structure group. The space-time is assumed to be a noncompact 
four-dimensional differentiable manifold having a countable base. The 
bundle ${\cal P}$ admits a connection $\Gamma $, whose translational 
and rotational parts of the coefficients will be written as 
$A^{k}{}_{\mu }$ and $A^{k}{}_{l\mu }$, respectively. The fundamental 
field variables are $A^{k}{}_{\mu }\;,A^{k}{}_{l\mu }$, the 
Higgs-type field $\psi=\{\psi^{k}\}$, and the matter field 
$\phi=\{\phi^{A}|A=1,2,3,\dots ,N\}$\cite{index}.
These fields transform according to\cite{derivative}
\begin{eqnarray}
\label{eq:transformation}
\psi^{\prime k}&=&(\Lambda(a^{-1}))^{k}{}_{l}(\psi^{l}-t^{l})\; ,
\nonumber \\
A^{\prime k}{}_{\mu }&=&
(\Lambda(a^{-1}))^{k}{}_{l}(A^{l}{}_{\mu }+t^{l}{}_{,\mu}
+A^{l}{}_{m\mu }t^{m})\; ,\nonumber \\
A^{\prime k}{}_{l\mu }&=&(\Lambda(a^{-1}))^{k}{}_{m}A^{m}{}_{n\mu }
(\Lambda(a))^{n}{}_{l}\nonumber \\
& &+(\Lambda(a^{-1}))^{k}{}_{m}(\Lambda(a))^{m}{}_{l,\mu}\; ,
\nonumber \\
\phi^{\prime A}&=&(\rho((t,a)^{-1}))^{A}{}_{B}\phi^{B}\; ,
\end{eqnarray}
under the $\overline{\mbox{Poincar\'e}}$ gauge transformation
\begin{eqnarray}
\label{eq:P-trans}
\sigma^{\prime }(x)
&=&\sigma(x)\cdot (t(x),a(x))\; ,\; \;\nonumber \\
& &t(x)\in T^{4}\; ,\; \; 
a(x)\in SL(2,C)\; .
\end{eqnarray}
Here, $\Lambda $ is the covering map from $SL(2,C)$ to the proper 
orthochronous Lorentz group, and $\rho $ stands for the 
representation of the $\overline{\mbox{Poincar\'e}}$ group to which 
the field $\phi^{A}$ belongs. Also, $\sigma $ and 
$\sigma^{\prime }$ stand for local cross sections of ${\cal P}$. 
Dual components $e^{k}{}_{\mu}$ of vierbeins 
$e^{\mu }{}_{k}\partial /\partial x^{\mu }$ are related to the
field $\psi^{k}$ and the gauge potentials $A^{k}{}_{\mu }$ and 
$A^{k}{}_{l\mu }$ through the relation
\begin{equation}
\label{eq:vierbein-comp}
e^{k}{}_{\mu}=\psi^{k}{}_{,\mu }+A^{k}{}_{l\mu}\psi^{l}
+A^{k}{}_{\mu}
\; ,
\end{equation}
and these transform according to
\begin{equation}
\label{eq:vierbein-trans}
e^{\prime k}{}_{\mu }=(\Lambda(a^{-1}))^{k}{}_{l}e^{l}{}_{\mu }\; ,
\end{equation}
under the transformation (\ref{eq:P-trans}). Also, they are related 
to the metric $g_{\mu \nu }dx^{\mu }\otimes dx^{\nu }$ of $M$ through 
the relation 
\begin{equation}
\label{eq:metric-vierbeins}
g_{\mu \nu }=\eta_{kl}e^{k}{}_{\mu }e^{l}{}_{\nu }
\end{equation}
with $(\eta_{kl})\stackrel{\mbox{\scriptsize def}}{=}
{\mbox{diag}}(-1,1,1,1)$.

There is a 2 to 1 bundle homomorphism $F$ from ${\cal P}$ to affine 
frame bundle ${\cal A}(M)$ over $M$, and an extended spinor structure 
and a spinor structure exist associated with it\cite{Kawai03}. The 
space-time $M$ is orientable, which follows from its assumed 
noncompactness and from the fact that $M$ has a spinor structure.   

The affine frame bundle ${\cal A}(M)$ admits a connection $\Gamma_{A}$. 
The $T^{4}$-part $\Gamma^{\mu }{}_{\nu }$ and $GL(4,R)$-part 
$\Gamma^{\mu }{}_{\nu \lambda }$ of its connection coefficients are 
related to $A^{k}{}_{l\mu }$ and $e^{k}{}_{\mu }$ through the 
relations
\begin{equation}
\label{eq:connection}
\Gamma^{\mu }{}_{\nu}=\delta^{\mu}{}_{\nu}\; ,\; \; \; 
A^{k}{}_{l\mu}=e^{k}{}_{\nu }e^{\lambda}{}_{l}
\Gamma^{\nu }{}_{\lambda \mu }+e^{k}{}_{\nu }e^{\nu }{}_{l,\mu }\; ,
\end{equation}
by the requirement that $F$ maps the connection $\Gamma $ into 
$\Gamma_{A}$, and the space-time $M$ is of the Riemann-Cartan 
type.  

The field strengths $R^{k}{}_{l\mu \nu }$, 
$R^{k}{}_{\mu\nu}$ and $T^{k}{}_{\mu \nu }$ of $A^{k}{}_{l\mu }$,
$A^{k}{}_{\mu}$ and of $e^{k}{}_{\mu}$ are given by
\cite{symmetrization-antisymmetrization}
\begin{eqnarray}
\label{eq:field-strength}
R^{k}{}_{l\mu \nu }&\stackrel{\mbox{\scriptsize def}}{=}&
2(A^{k}{}_{l[\nu ,\mu]}+A^{k}{}_{m[\mu}A^{m}{}_{l\nu]})\; ,
\nonumber \\
R^{k}{}_{\mu \nu }&\stackrel{\mbox{\scriptsize def}}{=}&
2(A^{k}{}_{[\nu ,\mu]}+A^{k}{}_{l[\mu}A^{l}{}_{\nu]})\; ,\nonumber \\ 
T^{k}{}_{\mu\nu}&\stackrel{\mbox{\scriptsize def}}{=}&
2(e^{k}{}_{[\nu ,\mu ]}+A^{k}{}_{l[\mu}e^{l}{}_{\nu]})\; ,
\end{eqnarray}
and we have the relation
\begin{equation}
\label{eq:field-strength-relation}
T^{k}{}_{\mu\nu}=R^{k}{}_{\mu\nu}+R^{k}{}_{l\mu\nu}\psi^{l}
\; .
\end{equation}
The field strengths $T^{k}{}_{\mu\nu}$ and $R^{kl}{}_{\mu\nu}$ are 
both invariant under \emph{internal\/} translations.
The torsion is given by 
\begin{equation}
\label{eq:torsion}
T^{\lambda}{}_{\mu \nu }=2\Gamma^{\lambda}{}_{[\nu \mu]}\; ,
\end{equation}
and the $T^{4}$- and $GL(4,R)$-parts of the curvature are given by
\begin{eqnarray}
\label{eq:curvature}
R^{\lambda}{}_{\mu \nu}&=&
2(\Gamma^{\lambda }{}_{[\nu ,\mu]}+\Gamma^{\lambda }{}_{\rho[\mu }
\Gamma^{\rho}{}_{\nu]})\; ,\\
R^{\lambda }{}_{\rho \mu \nu}&=&
2(\Gamma^{\lambda }{}_{\rho [\nu ,\mu]}+
\Gamma^{\lambda }{}_{\tau [\mu }
\Gamma^{\tau}{}_{\rho \nu]})\; ,
\end{eqnarray}
respectively.
Also, we have 
\begin{eqnarray}
\label{eq:strength-curvature-torsion}
T^{k}{}_{\mu \nu}&=&e^{k}{}_{\lambda }T^{\lambda }{}_{\mu \nu }=
e^{k}{}_{\lambda }R^{\lambda}{}_{\mu \nu }\; ,\\
R^{k}{}_{l\mu \nu }&=&e^{k}{}_{\lambda }e^{\rho}{}_{l}
R^{\lambda}{}_{\rho \mu \nu }\; ,
\end{eqnarray}
which follow from Eq. (\ref{eq:connection}).

The covariant derivative of the matter field $\phi $ takes the form
\begin{eqnarray}
\label{eq:cov-derivative matter fields}
D_{k}{\phi}^{A}&=&e^{\mu }{}_{k}D_{\mu }{\phi }^{A}\; ,\nonumber \\
D_{\mu }\phi^{A}&\stackrel{\mbox{\scriptsize def}}{=}&
\partial_{\mu}{\phi}^{A}+\frac{i}
{2}A^{lm}{}_{\mu}(M_{lm}{\phi})^{A}+iA^{l}{}_{\mu}(P_{l}{\phi})^{A}
\; .
\end{eqnarray}
Here, $M_{kl}$ and $P_{k}$ are representation matrices of the 
standard basis of the group $\bar{P}_{0}$ : 
$M_{kl}=-i\rho_{*}(\bar{M}_{kl})\;, P_{k}=
-i\rho_{*}(\bar{P}_{k})$. The matrix $P_{k}$ represents the intrinsic 
energy-momentum of the field $\phi^{A}$\cite{Kawai03}, and it is 
vanishing for all the observed fields.

The Lagrangian density\cite{raising-lowering}
\begin{equation}
\bar{L}=L^{M}(e^{k}{}_{\mu },\psi^{k},D_{k}\phi^{A},\phi^{A})
+\bar{L}^{G}(T^{klm},R^{klmn})
\end{equation}
satisfies the requirement of $\bar{P}_{0}$ gauge invariance. 
 Here, $L^{M}$ is the Lagrangian density of the matter field 
$\phi=\{\phi^{A}\}$ and $\bar{L}^{G}$ is the Lagrangian density of 
the gauge potentials given by 
\begin{equation}
\label{eq:lag-gauge}
\bar{L}^{G}\stackrel{\mbox{\scriptsize def}}{=}L^{T}+L^{R}+dR
\end{equation}
with
\widetext
\begin{eqnarray}
\label{eq:lag-T}
L^{T}&\stackrel{\mbox{\scriptsize def}}{=}&
c_{1}t^{klm}t_{klm}+c_{2}v^{k}v_{k}+c_{3}a^{k}a_{k}\; ,\\
\label{eq:lag-R}
L^{R}&\stackrel{\mbox{\scriptsize def}}{=}&
d_{1}A^{klmn}A_{klmn}+d_{2}B^{klmn}B_{klmn}+d_{3}C^{klmn}C_{klmn}
+d_{4}E^{kl}E_{kl}+d_{5}I^{kl}I_{kl}+d_{6}R^{2}\;.
\end{eqnarray} 
\narrowtext
\noindent In the above, $c_{i}, d_{j}$ ($i=1,2,3,j=1,2,3,\dots, 6$) 
and $d$ are  all real constants, $t_{klm},v_{k}$ and $a_{k}$ are the 
irreducible components of the field strength $T_{klm}$, and $A_{klmn}, 
B_{klmn},C_{klmn}, E_{kl},I_{kl}$, and $R$ are the irreducible components 
of the field strength $R_{klmn}$. Their definitions are enumerated in 
Appendix A.

The gravitational Lagrangian density $\bar{L}^{G}$ agrees with that 
in Poincar\'{e} gauge theory (P.G.T.)\cite{Hayashi-Shirafuji01}, and 
hence {\em gravitational field equations in} \=P.G.T. {\em take the same 
forms as those in} P.G.T. \cite{Newtonian-limit}.

In Refs.\cite{Kawai02,Kawai-Saitoh01,Kawai-Saitoh02}, we have used 
in place of $\bar{L}^{G}$,
\begin{equation}
\label{eq:grav-Lag}
L^{G}\stackrel{\mbox{\scriptsize def}}{=}
\bar{L}^{G}+{\mathbf \Delta}/\sqrt{-g}\; ,
\end{equation}
where we have defined 
\begin{equation}
\label{eq:g-Delta-def}
g\stackrel{\mbox{\scriptsize def}}{=}\det (g_{\mu \nu })\; ,\; \; \; 
{\mathbf \Delta}
\stackrel{\mbox{\scriptsize def}}{=}({\mathbf W}_{kl}{}^{\mu \nu }
A^{kl}{}_{\mu })_{,\nu }
\end{equation}
with 
\begin{equation}
\label{eq:W-def}
{\mathbf W}_{kl}{}^{\mu \nu }\stackrel{\mbox{\scriptsize def}}{=}
2d\sqrt{-g}e^{\mu }{}_{[k}e^{\nu }{}_{l]}\; .
\end{equation}
In order to get conserved generators, the Lagrangian density
$L\stackrel{\mbox{\scriptsize def}}{=}
L^{G}+L^{M}$, which leads to same field equations as $\bar{L}$ does, 
has been employed.

Let us consider $\overline{\mbox{Poincar\'e}}$ gauge transformations
(\ref{eq:P-trans}) with infinitesimal functions $t^{k}$ and with 
$a\in SL(2,C)$ such that $\Lambda(a)$ are
represented as
\begin{equation}
\label{eq:Lorentz-gauge-inf}
(\Lambda(a))^{k}{}_{l}=\delta^{k}{}_{l}+\omega^{k}{}_{l}
\end{equation} 
with infinitesimal functions 
$\omega_{kl}=-\omega_{lk}$. Also, we consider the infinitesimal 
coordinate transformations
\begin{equation}
\label{eq:coord-trans-inf}
x^{\prime \mu }=x^{\mu }+\epsilon^{\mu }
\end{equation}  
with $\epsilon^{\mu }$ being infinitesimal functions.
Under the product transformations of Eqs. (\ref{eq:Lorentz-gauge-inf})
and (\ref{eq:coord-trans-inf}), the fundamental fields 
$\psi^{k}, A^{k}{}_{\mu },A^{kl}{}_{\mu }$ and 
$\phi^{A}$ and the Lagrangian density $L$ transform according to
\begin{eqnarray}
\label{eq:transformation-inf}
\psi^{\prime k}&=&\psi^{k}-\omega^{k}{}_{l}\psi^{l}-t^{k}\; ,
\nonumber \\
A^{\prime k}{}_{\mu }&=&A^{k}{}_{\mu }-\omega^{k}{}_{l}A^{l}{}_{\mu }
+t^{k}{}_{,\mu }+A^{k}{}_{l\mu }t^{l}-
\epsilon^{\nu }{}_{,\mu }A^{k}{}_{\nu }\; ,\nonumber \\
A^{\prime kl}{}_{\mu }&=&A^{kl}{}_{\mu }
+\omega^{kl}{}_{,\mu }-\omega^{k}{}_{m}A^{ml}{}_{\mu }
-\omega^{l}{}_{m}A^{km}{}_{\mu }\nonumber \\
& &-\epsilon^{\nu }{}_{,\mu }A^{kl}{}_{\nu }\; ,
\nonumber \\
\phi^{\prime A}&=&\phi^{A}-it^{k}(P_{k}\phi)^{A}
-\frac{i}{2}\omega^{kl}(M_{kl}\phi)^{A}\; , \\
\label{eq:transformation-Lag}
L^{\prime }&=&L+\frac{1}{\sqrt{-g}}
\partial_{\nu }{\mathbf \Lambda}^{\nu }
\end{eqnarray}
with 
\begin{equation}
\label{eq:Lambda}
{\mathbf \Lambda}^{\nu }
\stackrel{\mbox{\scriptsize def}}{=}
\omega^{kl}{}_{,\mu }{\mathbf W}_{kl}{}^{\mu \nu }\; .  
\end{equation}
We see that $L$ is invariant under the product transformations of    
Eq. (\ref{eq:Lorentz-gauge-inf}) with {\em constant} $\omega_{kl}$
and Eq. (\ref{eq:coord-trans-inf}), but it  
violates {\em local} $SL(2,C)$ invariance.

In considering energy-momentum and angular momentum, there are two 
possibilities in choosing the set of independent field variables
\cite{Kawai02,Kawai-Saitoh01,Kawai-Saitoh02}, one is to choose the 
set $\{\psi^{k}, A^{k}{}_{\mu }, A^{kl}{}_{\mu },\phi^{A}\}$ and the 
other is to choose the set 
$\{\psi^{k}, e^{k}{}_{\mu },A^{kl}{}_{\mu },\phi^{A}\}$ instead. In 
the rest of this section, we employ 
$\{\psi^{k}, A^{k}{}_{\mu }, A^{kl}{}_{\mu },\phi^{A}\}$ as the set 
of independent field variables, because this choice is preferential 
to the other, as we have seen in Refs.
\cite{Kawai02,Kawai-Saitoh01,Kawai-Saitoh02}. The case when 
$\{\psi^{k}, e^{k}{}_{\mu },A^{kl}{}_{\mu },\phi^{A}\}$ is employed 
will be mentioned in the final section. 

From the transformation properties (\ref{eq:transformation-inf}) and 
(\ref{eq:transformation-Lag}), the identities\cite{Kawai01}
\begin{eqnarray}
\label{eq:identity-1}
& &\frac{\delta {\mathbf L}}{\delta \psi^{k}}
+\frac{\delta {\mathbf L}}{\delta A^{l}{}_{\mu }}A_{k}{}^{l}{}_{\mu }
+\left(\frac{\delta {\mathbf L}}{\delta A^{k}{}_{\mu }}\right)_{,\mu }
\nonumber \\
& &+i\frac{\delta {\mathbf L}}{\delta \phi^{A}}(P_{k}\phi )^{A}
\equiv 0\; ,\\
\label{eq:identity-2}
& &\frac{\delta {\mathbf L}}{\delta \psi^{[k}}\psi_{l]}
+\frac{\delta {\mathbf L}}{\delta A^{[k}{}_{\mu }}A_{l]\mu }
+\left(\frac{\delta {\mathbf L}}
{\delta A^{kl}{}_{\mu }}\right)_{,\mu }
\nonumber \\
& &+2\frac{\delta {\mathbf L}}
{\delta A^{[km}{}_{\mu }}A_{l]}{}^{m}{}_{\mu }
+\frac{i}{2}\frac{\delta {\mathbf L}}
{\delta \phi^{A}}(M_{kl}\phi )^{A}\equiv 0\; ,\\
\label{eq:identity-3}
& &{}^{\mbox{\rm tot}}{\mathbf T}_{k}{}^{\mu }
-{\mathbf F}_{k}{}^{\mu \nu }{}_{,\nu }
-\frac{\delta {\mathbf L}}{\delta A^{k}{}_{\mu }}\equiv 0\; ,\\
\label{eq:identity-4}
& &\left({}^{\mbox{\rm tot}}{\mathbf T}_{k}{}^{\mu }
-\frac{\delta {\mathbf L}}
{\delta A^{k}{}_{\mu }}\right)_{,\mu }\equiv 0\; ,\\
\label{eq:identity-5}
& &{}^{\mbox{\rm tot}}{\mathbf S}_{kl}{}^{\mu }
+2\frac{\delta {\mathbf L}}{\delta A^{kl}{}_{\mu }}
-{\mathbf \Sigma}_{kl}{}^{\mu \nu }{}_{,\nu }
\equiv 0\; ,\\
\label{eq:identity-6}
& &\left({}^{\mbox{\rm tot}}{\mathbf S}_{kl}{}^{\mu }
+2\frac{\delta {\mathbf L}}{\delta A^{kl}{}_{\mu }}\right)_{,\mu }   
\equiv 0\; ,\\
\label{eq:identity-9}
& &\tilde{\mathbf T}_{\mu }{}^{\nu }
-\partial_{\lambda }{\mathbf \Psi}_{\mu }{}^{\nu \lambda }
-\frac{\delta {\mathbf L}}{\delta A^{k}{}_{\nu }}
A^{k}{}_{\mu }
-\frac{\delta {\mathbf L}}{\delta A^{kl}{}_{\nu }}A^{kl}{}_{\mu }
\equiv 0
\end{eqnarray}
follow, where we have defined 
\begin{eqnarray}
\label{eq:def-1}
{\mathbf L}&\stackrel{\mbox{\scriptsize def}}{=}&\sqrt{-g}L\; ,\\
\label{eq:def-2}
{\mathbf F}_{k}{}^{\mu \nu }&\stackrel{\mbox{\scriptsize def}}{=}&   
\frac{\partial {\mathbf L}}{\partial A^{k}{}_{\mu,\nu }}
=\frac{\partial {\mathbf L}^{G}}{\partial A^{k}{}_{\mu,\nu }}=
-{\mathbf F}_{k}{}^{\nu \mu }\; ,\\
\label{eq:def-3}
{}^{\mbox{\rm tot}}{\mathbf T}_{k}{}^{\mu }
&\stackrel{\mbox{\scriptsize def}}{=}&
{\mathbf F}_{k}{}^{\mu }+i\frac{\partial {\mathbf L}}
{\partial \phi^{A}{}_{,\mu }}(P_{k}\phi)^{A}
+{\mathbf F}_{l}{}^{\nu \mu }A_{k}{}^{l}{}_{\nu}\; ,\\
\label{eq:def-4}
{}^{\mbox{\rm tot}}{\mathbf S}_{kl}{}^{\mu }
&\stackrel{\mbox{\scriptsize def}}{=}&
-2{\mathbf F}_{[k}{}^{\mu }\psi_{l]}
-2{\mathbf F}_{[k}{}^{\nu \mu}A_{l]\nu}\nonumber \\
& &-4{\mathbf F}_{[km}{}^{\nu \mu }A_{l]}{}^{m}{}_{\nu }
-i\frac{\partial {\mathbf L}}{\partial \phi^{A}{}_{,\mu }}
(M_{kl}\phi)^{A}\; ,\\
\label{eq:def-5}
\tilde{\mathbf T}_{\mu }{}^{\nu }&\stackrel{\mbox{\scriptsize def}}{=}&
\delta_{\mu }{}^{\nu }{\mathbf L}-{\mathbf F}_{k}{}^{\lambda \nu }
A^{k}{}_{\lambda ,\mu }
-{\mathbf F}_{kl}{}^{\lambda \nu }A^{kl}{}_{\lambda, \mu }\nonumber \\
& &-{\mathbf F}_{k}{}^{\nu }\psi^{k}{}_{,\mu }
-\frac{\partial {\mathbf L}}
{\partial \phi^{A}{}_{,\nu }}\phi^{A}{}_{,\mu }\; ,\\
\label{eq:def-6}
{\mathbf \Sigma}_{kl}{}^{\mu \nu }
&\stackrel{\mbox{\scriptsize def}}{=}&
-2{\mathbf F}_{kl}{}^{\mu \nu }
+2{\mathbf W}_{kl}{}^{\mu \nu }=-{\Sigma }_{kl}{}^{\nu \mu }
\end{eqnarray}
with 
\begin{eqnarray}
\label{eq:def-7}
{\mathbf F}_{k}{}^{\mu }&\stackrel{\mbox{\scriptsize def}}{=}&
\frac{\partial {\mathbf L}}{\partial \psi^{k}{}_{,\mu }}\; ,\\
\label{eq:def-8}
{\mathbf F}_{kl}{}^{\mu \nu }&\stackrel{\mbox{\scriptsize def}}{=}&
\frac{\partial {\mathbf L}}{\partial A^{kl}{}_{\mu,\nu }}
=\frac{\partial {\mathbf L}^{G}}{\partial A^{kl}{}_{\mu,\nu }}
=-{\mathbf F}_{kl}{}^{\nu \mu }\; ,\\
\label{eq:def-9}
{\mathbf \Psi}_{\lambda}{}^{\mu \nu }
&\stackrel{\mbox{\scriptsize def}}{=}&
{\mathbf F}_{k}{}^{\mu \nu }A^{k}{}_{\lambda}+
{\mathbf F}_{kl}{}^{\mu \nu }A^{kl}{}_{\lambda }
=-{\mathbf \Psi}_{\lambda}{}^{\nu \mu }\; . 
\end{eqnarray}
 
The energy-momentum density ${}^{\mbox{\rm tot}}
{\mathbf T}_{k}{}^{\mu }$ 
and the \lq \lq spin" angular momentum density 
${}^{\mbox{\rm tot}}{\mathbf S}_{kl}{}^{\mu }$ are   
expressed as follows:
\begin{eqnarray}
\label{eq:dyn-em-dens-alt}
{}^{\mbox{\rm tot}}{\mathbf T}_{k}{}^{\mu }
&=&\frac{\partial {\mathbf L}}{\partial A^{k}{}_{\mu }}\; ,\\
\label{eq:spin-dens-alt}
{}^{\mbox{\rm tot}}{\mathbf S}_{kl}{}^{\mu }
&=&-2\frac{\partial {\mathbf L}}{\partial A^{kl}{}_{\mu }}
+2{\mathbf W}_{kl}{}^{\mu \nu }{}_{,\nu }\; ,
\end{eqnarray}
by virtue of the identities (\ref{eq:identity-3}) and 
(\ref{eq:identity-5}). Thus, ${}^{\mbox{\rm tot}}
{\mathbf T}_{k}{}^{\mu }$ has the standard form of gauge current in 
Yang-Mills theories, while ${}^{\mbox{\rm tot}}
{\mathbf S}_{kl}{}^{\mu }$ has not. There is an additional term 
$2{\mathbf W}_{kl}{}^{\mu \nu }{}_{,\nu }$ which originates from the 
term ${\mathbf {\Delta }}$ violating the $SL(2,C)$-gauge invariance 
of the gravitational Lagrangian density.

When the field equations $\delta{\mathbf L}/\delta A^{k}{}_{\mu }
\stackrel{\mbox{\scriptsize def}}{=}
\partial{\mathbf L}/\partial A^{k}{}_{\mu}
-\partial_{\nu }(\partial{\mathbf L}/\partial A^{k}{}_{\mu ,\nu})=0$ 
and $\delta {\mathbf L}/\delta \phi^{A}=0$ are both satisfied, 
we have the following:\\
(i) The field equation $\delta {\mathbf L}/{\delta \psi^{k}}=0$ 
is automatically satisfied, and hence $\psi^{k}$ is not an 
independent dynamical variable.\\
(ii) 
\begin{eqnarray}
\label{eq:diff-consv-1}
\partial_{\mu }{}^{\mbox{\rm tot}}{\mathbf T}_{k}{}^{\mu }=0\; ,\\
\label{eq:diff-consv-2}
\partial_{\mu }{}^{\mbox{\rm tot}}{\mathbf S}_{kl}{}^{\mu }=0\; ,
\end{eqnarray}
which are the differential conservation laws of the dynamical 
energy-momentum and of the \lq \lq spin" angular momentum, 
respectively. (i) and (ii) follow from 
Eqs. (\ref{eq:identity-1}), (\ref{eq:identity-4}), and 
(\ref{eq:identity-6}).

Equations (\ref{eq:identity-9}) and (\ref{eq:def-9}) 
lead to 
\begin{eqnarray}
\label{eq:diff-consv-3}
\partial_{\nu }\tilde{\mathbf T}_{\mu }{}^{\nu }&=&0\; ,\\
\label{eq:diff-consv-4}
\partial_{\nu}\tilde{\mathbf M}_{\lambda}{}^{\mu \nu }&=&0\; ,
\end{eqnarray}
when $\delta {\mathbf L}/\delta A^{k}{}_{\mu }=0\, ,
\delta {\mathbf L}/\delta A^{kl}{}_{\mu }=0$, where 
$\tilde{\mathbf M}_{\lambda }{}^{\mu \nu }
\stackrel{\mbox{\scriptsize def}}{=}
2({\mathbf \Psi }_{\lambda }{}^{\mu \nu }-x^{\mu }
\tilde{\mathbf T}_{\lambda}{}^{\nu })$. Equations 
(\ref{eq:diff-consv-3}) and (\ref{eq:diff-consv-4}) are the 
differential conservation laws of the canonical energy-momentum and 
\lq \lq extended orbital angular momentum"
\cite{Kawai-Saitoh01,Kawai-Saitoh02}, respectively.

In Refs.\cite{Kawai02,Kawai-Saitoh01,Kawai-Saitoh02}, we have 
examined the integrations of 
${}^{\mbox{\rm tot}}{\mathbf T}_{k}{}^{\mu }, 
{}^{\mbox{\rm tot}}{\mathbf S}_{kl}{}^{\mu }, 
\tilde{\mathbf T}_{\mu }{}^{\nu }$ and 
$\tilde{\mathbf M}_{\lambda }{}^{\mu \nu }$
for asymptotically flat space-time by choosing $\psi^{k}$ as
\begin{eqnarray}
\label{eq:psi-asymp}
\psi^{k}&=&e^{(0)k}{}_{\mu }x^{\mu }+\psi^{(0)k}+O(1/r^{\beta})\; ,
\nonumber \\
\psi^{k}{}_{,\mu }&=&e^{(0)k}{}_{\mu }+O(1/r^{\beta +1})\; ,\; \; \; 
(\beta>0)\; ,
\end{eqnarray}
where $e^{(0)k}{}_{\mu }$ is a constant 
satisfying $e^{(0)k}{}_{\mu }\eta_{kl}e^{(0)l}{}_{\nu }
=\eta_{\mu \nu}$,
and $\psi^{(0)k}$ and $\beta$ are constants. Also, we have defined 
$r\stackrel{\mbox{\scriptsize def}}{=}
\sqrt{(x^{1})^{2}+(x^{2})^{2}+(x^{3})^{2}}$, and $O(1/r^{\alpha})$ 
with positive $\alpha $ denotes a term for which 
$r^{\alpha }O(1/r^{\alpha })$ remains finite for 
$r \rightarrow \infty $; a term $O(1/r^{\alpha })$ may of course 
be zero. We have shown the following:
\begin{eqnarray}
\label{eq:em}  
M_{k}&\stackrel{\mbox{\scriptsize def}}{=}&
\int_{\sigma}{}^{\mbox{\rm tot}}{\mathbf T}_{k}{}^{\mu }
d\sigma_{\mu }
=e^{(0)\mu }{}_{k}M_{\mu }\; ,\\
\label{eq:ang}
S_{kl}&\stackrel{\mbox{\scriptsize def}}{=}&
\int_{\sigma}{}^{\mbox{\rm tot}}{\mathbf S}_{kl}{}^{\mu }
d\sigma_{\mu }
=e^{(0)}{}_{k\mu }e^{(0)}{}_{l\nu }M^{\mu \nu }\nonumber \\
& &+2\psi^{(0)}{}_{[k}M_{l]}\; , \\
\label{eq:can-em}
M^{c}{}_{\mu }&\stackrel{\mbox{\scriptsize def}}{=}&
\int_{\sigma}\tilde{\mathbf T}_{\mu }{}^{\nu }d\sigma_{\nu }=0\; ,\\
\label{eq:ext-orb-ang}
L_{\mu }{}^{\nu }
&\stackrel{\mbox{\scriptsize def}}{=}&
\int_{\sigma}\tilde{\mathbf M}_{\mu}{}^{\nu \lambda}
d\sigma_{\lambda }=0\; ,
\end{eqnarray}
where $d\sigma_{\mu }$ denotes the surface element on a space-like 
surface $\sigma$. Also, we have defined
\begin{eqnarray}
\label{eq:Mmu}
M_{\mu }&\stackrel{\mbox{\scriptsize def}}{=}&
\eta_{\mu \lambda}\int_{\sigma }\theta^{\lambda \nu }
d\sigma_{\nu }\; ,\\
\label{eq:Mmunu}
M^{\mu \nu }&\stackrel{\mbox{\scriptsize def}}{=}&
\int_{\sigma }\partial_{\rho}K^{\mu \nu \lambda \rho }
d\sigma_{\lambda}\nonumber \\
&=&\int_{\sigma}(x^{\mu }\theta^{\nu \lambda}-x^{\nu }
\theta^{\mu \lambda})
d\sigma_{\lambda}
\end{eqnarray}
with 
\begin{eqnarray}
\label{eq:theta}
\theta^{\lambda \nu }&\stackrel{\mbox{\scriptsize def}}{=}&
2d\partial_{\rho}\partial_{\sigma}
\{(-g)g^{\lambda [\nu}g^{\rho ]\sigma }\}\; ,\\
\label{eq:K}
K^{\mu \nu \lambda \rho}&\stackrel{\mbox{\scriptsize def}}{=}&
2d[x^{\mu }\partial_{\sigma }\{(-g)g^{\nu [\lambda}g^{\rho ]\sigma}\}
-x^{\nu }\partial_{\sigma }\{(-g)g^{\mu [\lambda}g^{\rho ]\sigma}\}
\nonumber \\
& &+(-g)g^{\mu [\lambda}g^{\nu \rho ]}]\; .
\end{eqnarray}
Actually, Eq. (\ref{eq:em}) has been obtained without using 
Eq. (\ref{eq:psi-asymp}), but it is crucial in 
obtaining the expressions (\ref{eq:ang}) $\sim $ 
(\ref{eq:ext-orb-ang}). Also, the expression of 
$\theta^{\lambda \nu }$ agrees with that of the symmetric 
energy-momentum density proposed by Landau-Lifshitz in general 
relativity.
 
The dynamical energy-momentum $M_{k}$ is the generator of internal 
translations
 and the total energy-momentum of the system. The \lq \lq spin" 
angular momentum $S_{kl}$ is the generator of internal $SL(2,C)$-
transformations and the {\em total} (={\em spin}+{\em orbital}) 
angular momentum of the system.
The canonical energy-momentum $M^{c}{}_{\mu }$ and the \lq \lq 
extended orbital angular momentum" $L_{\mu }{}^{\nu }$ are the 
generators of coordinate translations and of coordinate $GL(4,R)$- 
transformations, respectively\cite{Kawai-Saitoh02,extended}.

The following is worth emphasizing: The total energy-momentum and 
the total angular momentum are the generators of {\em internal} 
$\overline{\mbox{\rm Poincar\'{e}}}$ transformations, and the 
generators of coordinate transformations are vanishing and trivial. 
\subsection{Transformation properties of energy-momentum and angular 
momentum densities}
\label{subsubsec-trans-PGT}
We define densities ${}^{G}{\mathbf T}_{k}{}^{\mu },
{}^{M}{\mathbf T}_{k}{}^{\mu }, 
{}^{G}{\mathbf S}_{kl}{}^{\mu }$ and 
${}^{M}{\mathbf S}_{kl}{}^{\mu }$ by 
\begin{eqnarray}
\label{eq:GM-em-spin}
{}^{G}{\mathbf T}_{k}{}^{\mu }
&\stackrel{\mbox{\scriptsize def}}{=}&
\frac{\partial {\mathbf L}^{G}}{\partial\psi^{k}{}_{,\mu }}+
{\mathbf F}_{l}{}^{\nu \mu }A_{k}{}^{l}{}_{\nu }
=\frac{\partial{\mathbf L}^{G}}{\partial A^{k}{}_{\mu }}\; ,\\ 
 {}^{M}{\mathbf T}_{k}{}^{\mu }
&\stackrel{\mbox{\scriptsize def}}{=}&
\frac{\partial {\mathbf L}^{M}}{\partial \psi^{k}{}_{,\mu }}
+i\frac{\partial{\mathbf L}^{M}}
{\partial \phi^{A}{}_{,\mu }}(P_{k}\phi)^{A}
=\frac{\partial{\mathbf L}^{M}}{\partial A^{k}{}_{\mu }}\; ,\\ 
{}^{G}{\mathbf S}_{kl}{}^{\mu }
&\stackrel{\mbox{\scriptsize def}}{=}&
-2\frac{\partial {\mathbf L}^{G}}
{\partial \psi^{[k}{}_{,\mu }}\psi_{l]}
-2{\mathbf F}_{[k}{}^{\nu \mu }A_{l]\nu }
-4{\mathbf F}_{[km}{}^{\nu \mu }A_{l]}{}^{m}{}_{\nu }\nonumber \\
&=&-2\frac{\partial{\mathbf L}^{G}}{\partial A^{kl}{}_{\mu }} 
+2{\mathbf W}_{kl}{}^{\mu \nu }{}_{,\nu }\; ,\\
{}^{M}{\mathbf S}_{kl}{}^{\mu }
&\stackrel{\mbox{\scriptsize def}}{=}&
-2\frac{\partial {\mathbf L}^{M}}
{\partial \psi^{[k}{}_{,\mu }}\psi_{l]}
-i\frac{\partial {\mathbf L}^{M}}
{\partial \phi^{A}{}_{,\mu }}(M_{kl}\phi)^{A}
\nonumber \\
& &=-2\frac{\partial{\mathbf L}^{M}}
{\partial A^{kl}{}_{\mu }}\; 
\end{eqnarray}
with ${\mathbf L}^{G}\stackrel{\mbox{\scriptsize def}}{=}
\sqrt{-g}L^{G}$ and
${\mathbf L}^{M}\stackrel{\mbox{\scriptsize def}}{=}\sqrt{-g}L^{M}$. 
The densities ${}^{G}{\mathbf T}_{k}{}^{\mu }$ and 
${}^{M}{\mathbf T}_{k}{}^{\mu }$
are the dynamical energy-momentum densities of the gravitational 
field and of the matter field $\phi^{A}$, respectively, while 
${}^{G}{\mathbf S}_{kl}{}^{\mu }$ and 
${}^{M}{\mathbf S}_{kl}{}^{\mu }$ are \lq \lq spin" angular momentum 
densities of the gravitational and the matter fields, respectively. 
There are 
the relations
\begin{equation}
\label{eq:tot=G+M}
{}^{\mbox{\rm tot}}{\mathbf T}_{k}{}^{\mu }=
{}^{G}{\mathbf T}_{k}{}^{\mu }+{}^{M}{\mathbf T}_{k}{}^{\mu }\; ,
\; \; \; {}^{\mbox{\rm tot}}{\mathbf S}_{kl}{}^{\mu }=
{}^{G}{\mathbf S}_{kl}{}^{\mu }+{}^{M}{\mathbf S}_{kl}{}^{\mu }\; . 
\end{equation}
Under the product transformations of Eqs. (\ref{eq:Lorentz-gauge-inf})
and (\ref{eq:coord-trans-inf}), the densities 
${}^{G}{\mathbf T}_{k}{}^{\mu },{}^{M}{\mathbf T}_{k}{}^{\mu },$
${}^{G}{\mathbf S}_{kl}{}^{\mu }$,
and ${}^{M}{\mathbf S}_{kl}{}^{\mu }$ transform according to
\widetext
\begin{eqnarray}
\label{eq:em-G-trans-PGT}
{}^{G}{\mathbf T}^{\prime }{}_{k}{}^{\mu }
&=&{}^{G}{\mathbf T}_{k}{}^{\mu }
-\omega_{k}{}^{l}{}^{G}{\mathbf T}_{l}{}^{\mu }
+\epsilon^{\mu}{}_{,\nu }{}^{G}{\mathbf T}_{k}{}^{\nu }
-\epsilon^{\lambda}{}_{,\lambda }{}^{G}{\mathbf T}_{k}{}^{\mu }-
\omega_{k}{}^{l}{}_{,\nu }{\mathbf F}_{l}{}^{\mu \nu }+
\frac{\partial (\partial_{\nu }{\mathbf \Lambda }^{\nu })}
{\partial A^{k}{}_{\mu }}\; ,\\
\label{eq:em-M-trans-PGT}
{}^{M}{\mathbf T}^{\prime }{}_{k}{}^{\mu }
&=&
{}^{M}{\mathbf T}_{k}{}^{\mu }
-\omega_{k}{}^{l}{}^{M}{\mathbf T}_{l}{}^{\mu }
+\epsilon^{\mu}{}_{,\nu }{}^{M}{\mathbf T}_{k}{}^{\nu }
-\epsilon^{\lambda}{}_{,\lambda }{}^{M}{\mathbf T}_{k}{}^{\mu }\; ,\\
\label{eq:spin-G-trans-PGT}
{}^{G}{\mathbf S}^{\prime }{}_{kl}{}^{\mu }
&=&{}^{G}{\mathbf S}{}_{kl}{}^{\mu }
-\omega_{k}{}^{m}{}^{G}{\mathbf S}{}_{ml}{}^{\mu }
-\omega_{l}{}^{m}{}^{G}{\mathbf S}{}_{km}{}^{\mu }
-2t_{[k}{}^{G}{\mathbf T}{}_{l]}{}^{\mu }
+\epsilon^{\mu }{}_{,\nu }{}^{G}{\mathbf S}{}_{kl}{}^{\nu }
-\epsilon^{\lambda}{}_{,\lambda}
{}^{G}{\mathbf S}{}_{kl}{}^{\mu }\nonumber \\
& &-2t_{[k,\nu }{\mathbf F}_{l]}{}^{\mu \nu }
-2\omega_{[k}{}^{m}{}_{,\nu }{\mathbf \Sigma}_{l]m}{}^{\mu \nu }
-2\frac{\partial (\partial_{\nu }{\mathbf \Lambda }^{\nu })}
{\partial A^{kl}{}_{\mu }}\; ,\\
\label{eq:spin-M-trans-PGT}
{}^{M}{\mathbf S}^{\prime }{}_{kl}{}^{\mu }
&=&{}^{M}{\mathbf S}{}_{kl}{}^{\mu }
-\omega_{k}{}^{m}{}^{M}{\mathbf S}{}_{ml}{}^{\mu }
-\omega_{l}{}^{m}{}^{M}{\mathbf S}{}_{km}{}^{\mu }
-2t_{[k}{}^{M}{\mathbf T}{}_{l]}{}^{\mu }
+\epsilon^{\mu }{}_{,\nu }{}^{M}{\mathbf S}{}_{kl}{}^{\nu }
-\epsilon^{\lambda}{}_{,\lambda}
{}^{M}{\mathbf S}{}_{kl}{}^{\mu }\; .
\end{eqnarray}
\narrowtext
We define ${}^{G}\tilde{\mathbf T}_{\mu }{}^{\nu }\; ,
{}^{M}{\mathbf T}_{\mu }{}^{\nu }$ by
\begin{eqnarray}
\label{eq:canonical-grav}
{}^{G}\tilde{\mathbf T}_{\mu }{}^{\nu }
&\stackrel{\mbox{\scriptsize def}}{=}&
\delta_{\mu }{}^{\nu }{\mathbf L}^{G}-
{\mathbf F}_{k}{}^{\lambda \nu }
A^{k}{}_{\lambda ,\mu }
-{\mathbf F}_{kl}{}^{\lambda \nu }A^{kl}{}_{\lambda, \mu }
\nonumber \\
& &-\frac{\partial {\mathbf L}^{G}}{\partial \psi^{k}{}_{,\nu }}
\psi^{k}{}_{,\mu }\;, \\
\label{eq:canonical-matter}
{}^{M}{\mathbf T}_{\mu }{}^{\nu }
&\stackrel{\mbox{\scriptsize def}}{=}&
\delta_{\mu }{}^{\nu }{\mathbf L}^{M}
-\frac{\partial {\mathbf L}^{M}}{\partial \psi^{k}{}_{,\nu}}
\psi^{k}{}_{,\mu }-\frac{\partial {\mathbf L}^{M}}
{\partial \phi^{A}{}_{,\nu }}\phi^{A}{}_{,\mu }\; ,
\end{eqnarray}
which are the canonical energy-momentum densities of the 
gravitational field and of the matter field, respectively. 
Also, we define
\begin{eqnarray}
\label{eq:extended-orbital-grav}
{}^{G}\tilde{\mathbf M}{}_{\lambda}{}^{\mu \nu }
&\stackrel{\mbox{\scriptsize def}}{=}&
2({\mathbf \Psi}_{\lambda}{}^{\mu \nu }-x^{\mu}{}^{G}
\tilde{\mathbf T}_{\lambda}{}^{\nu })\; ,\\
\label{eq:extended-orbital-matter}
{}^{M}\tilde{\mathbf M}{}_{\lambda}{}^{\mu \nu }
&\stackrel{\mbox{\scriptsize def}}{=}&
-2x^{\mu }{}^{M}{\mathbf T}_{\lambda}{}^{\nu }\; ,
\end{eqnarray}
which are the \lq \lq extended orbital angular momentum" densities 
of the gravitational field and of the matter field, respectively.
There are the relations
\begin{equation}
\label{eq;tot=G+M}
\tilde{\mathbf T}_{\mu}{}^{\nu }
={}^{G}\tilde{\mathbf T}_{\mu }{}^{\nu }+
{}^{M}{\mathbf T}_{\mu }{}^{\nu }\; ,\;  \; \;
\tilde{\mathbf M}_{\lambda}{}^{\mu \nu}
={}^{G}\tilde{\mathbf M}_{\lambda}{}^{\mu \nu }+
{}^{M}\tilde{\mathbf M}_{\lambda}{}^{\mu \nu }\; .
\end{equation}
The densities 
${}^{G}\tilde{\mathbf T}_{\mu }{}^{\nu },
{}^{M}{\mathbf T}_{\mu }{}^{\nu },
{}^{G}\tilde{\mathbf M}_{\lambda}{}^{\mu \nu }$ and 
${}^{M}\tilde{\mathbf M}_{\lambda}{}^{\mu \nu }$ transform according 
to
\widetext
\begin{eqnarray}
\label{eq:can-em-G-trans}
{}^{G}\tilde {\mathbf T}^{\prime }{}_{\mu }{}^{\nu }
&=&{}^{G}\tilde {\mathbf T}{}_{\mu }{}^{\nu }-
(t^{k}{}_{,\lambda \mu }
+A^{k}{}_{l\lambda }t^{l}{}_{,\mu })
{\mathbf F}_{k}{}^{\lambda \nu }
-(\omega^{kl}{}_{,\lambda \mu }
-\omega^{k}{}_{m,\mu }A^{ml}{}_{\lambda }
-\omega^{l}{}_{m,\mu }A^{km}{}_{\lambda })
{\mathbf F}_{kl}{}^{\lambda \nu }\nonumber \\
& &+t^{k}{}_{,\mu }
\frac{\partial {\mathbf L}^{G}}{\partial \psi^{k}{}_{,\nu }}
+\omega^{kl}{}_{,\mu}
\frac{\partial {\mathbf L}^{G}}{\partial \psi^{[k}{}_{,\nu }}
\psi_{l]}
+(\partial_{\rho}{\mathbf \Lambda}^{\rho})
\delta_{\mu }{}^{\nu }
-\frac{\partial(\partial_{\rho}{\mathbf \Lambda}^{\rho})}
{\partial \psi^{k}{}_{,\nu }}\psi^{k}{}_{,\mu }
-\frac{\partial(\partial_{\rho}{\mathbf \Lambda}^{\rho})}
{\partial A^{k}{}_{\lambda ,\nu }}
A^{k}{}_{\lambda ,\mu }\nonumber \\
& &-\frac{\partial (\partial_{\rho}{\mathbf \Lambda}^{\rho})}
{\partial A^{kl}{}_{\lambda ,\nu }}
A^{kl}{}_{\lambda ,\mu }
+\epsilon^{\lambda}{}_{,\mu }
{}^{G}\tilde{\mathbf T}_{\lambda }{}^{\nu }
-\epsilon^{\nu }{}_{,\lambda }{}^{G}
\tilde{\mathbf T}_{\mu }{}^{\lambda }
-\epsilon^{\lambda }{}_{,\lambda }{}^{G}
\tilde{\mathbf T}_{\mu }{}^{\nu }
+\epsilon^{\rho}{}_{,\lambda \mu }
{\mathbf \Psi}_{\rho}{}^{\lambda \nu}\; ,\\
\label{eq:can-em-M-trans}
{}^{M}{\mathbf T}^{\prime }{}_{\mu }{}^{\nu }
&=&{}^{M}{\mathbf T}{}_{\mu }{}^{\nu }
+t^{k}{}_{,\mu }\frac{\partial {\mathbf L}^{M}}
{\partial \psi^{k}{}_{,\nu }}
+\omega^{kl}{}_{,\mu}
\frac{\partial {\mathbf L}^{M}}{\partial \psi^{[k}{}_{,\nu }}
\psi_{l]}+\epsilon^{\lambda}{}_{,\mu }
{}^{M}{\mathbf T}_{\lambda }{}^{\nu }
-\epsilon^{\nu }{}_{,\lambda }{}^{M}
{\mathbf T}_{\mu }{}^{\lambda }
-\epsilon^{\lambda }{}_{,\lambda }{}^{M}{\mathbf T}_{\mu }{}^{\nu }
\; , \\
\label{eq:ext-orb-G-trans}
{}^{G}\tilde{\mathbf M}^{\prime }{}_{\lambda}{}^{\mu \nu }
&=&2({\mathbf \Psi}^{\prime }{}_{\lambda }{}^{\mu \nu } 
-x^{\mu } {}^{G}\tilde {\mathbf T}^{\prime }{}_{\lambda }{}^{\nu }
-\epsilon^{\mu }{}^{G}\tilde{\mathbf T}{}_{\lambda }{}^{\nu })\; ,\\
\label{eq:ext-orb-M-trans}
{}^{M}\tilde{\mathbf M}^{\prime }{}_{\lambda}{}^{\mu \nu }
&=&-2x^{\mu } {}^{M}{\mathbf T}^{\prime }{}_{\lambda }{}^{\nu }
-2\epsilon^{\mu }{}^{M}{\mathbf T}{}_{\lambda }{}^{\nu }\; ,
\end{eqnarray}
under the product transformations of Eqs. (\ref{eq:Lorentz-gauge-inf})
and (\ref{eq:coord-trans-inf}), where 
${\mathbf \Psi}{}^{\prime }{}_{\lambda }{}^{\mu \nu }$
 denotes the transformed ${\mathbf \Psi}{}_{\lambda }{}^{\mu \nu }$:
\begin{eqnarray}
\label{eq:Psi-trans}
{\mathbf \Psi}{}^{\prime }{}_{\lambda }{}^{\mu \nu }
&=&{\mathbf \Psi}{}_{\lambda }{}^{\mu \nu }+
t^{k}{}_{,\lambda }{\mathbf F}_{k}{}^{\mu \nu }
+\omega^{kl}{}_{,\lambda }{\mathbf F}_{kl}{}^{\mu \nu }
+\frac{\partial (\partial_{\rho}{\Lambda }^{\rho})}
{\partial A^{k}{}_{\mu ,\nu }}A^{k}{}_{\lambda }
+\frac{\partial (\partial_{\rho}{\Lambda }^{\rho})}
{\partial A^{kl}{}_{\mu ,\nu }}A^{kl}{}_{\lambda }-
\epsilon^{\rho}{}_{,\lambda }{\mathbf \Psi}{}_{\rho }{}^{\mu \nu }
+\epsilon^{\mu}{}_{,\rho}{\mathbf \Psi}{}_{\lambda }{}^{\rho \nu }
\nonumber \\
& &+\epsilon^{\nu}{}_{,\rho}{\mathbf \Psi}{}_{\lambda }{}^{\mu \rho }
-\epsilon^{\rho}{}_{,\rho}{\mathbf \Psi}{}_{\lambda }{}^{\mu \nu }\; .
\end{eqnarray}
\narrowtext
\section{EXTENDED NEW GENERAL RELATIVITY}
\label{sec:ENGR}
\subsection{Reduction of $\overline{\mbox{Poincar\'e}}$~gauge
theory to extended new general relativity}
\label{subsec:reduction}

In \={P}.G.T., we consider the case in which the field strength 
$R^{kl}{}_{\mu \nu }$ vanishes identically,
\begin{equation}
\label{eq:field-strength-vanish}
R^{kl}{}_{\mu \nu}\equiv0 \; ,
\end{equation}
then, the curvature vanishes and we have a teleparallel theory. 

We choose the $SL(2,C)$-gauge such that
\begin{equation}
\label{eq:gauge-condition}
A^{kl}{}_{\mu}\equiv 0\; ,
\end{equation}
which reduces the expressions of vierbeins $e^{k}{}_{\mu}$, affine 
connection coefficients ${\Gamma}^{\lambda}{}_{\mu \nu}$ and the 
covariant derivative $D_{k}\phi$ to   
\begin{eqnarray}
\label{eq:vierbein-ENGR}
e^{k}{}_{\mu}&=&{\psi}^{k}{}_{,\mu}+A^{k}{}_{\mu}\; ,\\
\label{eq:affine-connec}
{\Gamma}^{\lambda}{}_{\mu\nu}&=&
e^{\lambda}{}_{l}\,e^{l}{}_{\mu,\nu}\; ,\\
\label{eq:cov-deri-ENGR}
D_{k}\phi^{A}&=&e^{\mu }{}_{k}D_{\mu}\phi^{A}\; ,\\
D_{\mu }\phi^{A}&=&\partial_{\mu }\phi^{A}+
iA^{l}{}_{\mu }(P_{l}\phi)^{A}
\; ,
\end{eqnarray}
respectively.

Since $L^{R}=0=dR$, the gravitational Lagrangian density $L^{G}$ is 
reduced to $L^{G}=L^{T}$, which agrees with the gravitational 
Lagrangian density in new general relativity (N.G.R.)
\cite{Hayashi-Shirafuji02,probable-parameters}. Thus, {\em the 
gravitational field equations in} E.N.G.R. {\em take the same 
forms as those in} N.G.R.

As is the case of the Lagrangian density $L^{G}+L^{M}$ in \={P}.G.T., 
$L=L^{T}+L^{M}$ is invariant under the product transformations 
of Eq. (\ref{eq:Lorentz-gauge-inf}) with {\em constant} $\omega_{kl}$ and  
Eq. (\ref{eq:coord-trans-inf}), but it violates {\em local} $SL(2,C)$ 
invariance. 

The identities (\ref{eq:identity-1}) and (\ref{eq:identity-2}) and the 
definitions (\ref{eq:def-3}) and (\ref{eq:def-4}) are reduced to
\cite{independent-set}
\begin{eqnarray}
\label{eq:identity-ENGR1}
& &\frac{\delta {\mathbf L}}{\delta \psi^{k}}
+\left(\frac{\delta {\mathbf L}}
{\delta A^{k}{}_{\mu }}\right)_{,\mu }
+i\frac{\delta {\mathbf L}}{\delta \phi^{A}}(P_{k}\phi )^{A}
\equiv 0\;, \\
\label{eq:identity-ENGR2}
& &\partial_{\mu }{}^{\mbox{\rm tot}}{\mathbf S}_{kl}{}^{\mu }
-2\frac{\delta {\mathbf L}}{\delta \psi^{[k}}\psi_{l]}
-2\frac{\delta {\mathbf L}}{\delta A^{[k}{}_{\mu }}A_{l]\mu}
-i\frac{\delta {\mathbf L}}{\delta \phi^{A}}(M_{kl}\phi)^{A}
\nonumber \\
& &\equiv 0\; ,\\
\label{eq:def-ENGR1}
& &{}^{\mbox{\rm tot}}{\mathbf T}_{k}{}^{\mu }
={\mathbf F}_{k}{}^{\mu }
+i\frac{\partial {\mathbf L}}
{\partial \phi^{A}{}_{,\mu }}(P_{k}\phi)^{A}
\; ,\\
\label{eq:def-ENGR2}
& &{}^{\mbox{\rm tot}}{\mathbf S}_{kl}{}^{\mu }
=-2{\mathbf F}_{[k}{}^{\mu }\psi_{l]}
-2{\mathbf F}_{[k}{}^{\nu \mu }A_{l]\nu }\nonumber \\
& &\mbox{\hspace*{10ex}}-i\frac{\partial {\mathbf L}}
{\partial \phi^{A}{}_{,\mu }}(M_{kl}\phi)^{A}\; ,
\end{eqnarray}
respectively. The identities (\ref{eq:identity-3}) and  
(\ref{eq:identity-4}), the expression (\ref{eq:dyn-em-dens-alt}), and 
the conservation laws (\ref{eq:diff-consv-1}) and (\ref{eq:diff-consv-2}) 
remain unchanged and there is no identity corresponding to 
Eqs. (\ref{eq:identity-5}) and (\ref{eq:identity-6}) and no expression corresponding to Eq. (\ref{eq:spin-dens-alt}).

The field equation $\delta {\mathbf L}/{\delta \psi^{k}}=0$ 
is automatically satisfied, if the field equations 
$\delta{\mathbf L}/\delta A^{k}{}_{\mu }=0$ and
$\delta {\mathbf L}/\delta \phi^{A}=0$ are both satisfied. 

The identity (\ref{eq:identity-9}), the definitions (\ref{eq:def-5}) 
and (\ref{eq:def-9}) are reduced to
\begin{eqnarray}
\label{eq:identity-ENGR5}
& &\tilde{\mathbf T}_{\mu }{}^{\nu }-
\partial_{\lambda }{\mathbf \Psi}_{\mu }{}^{\nu \lambda }-
\frac{\delta {\mathbf L}}{\delta A^{k}{}_{\nu }}A^{k}{}_{\mu }
\equiv 0\; ,\\
\label{eq:def-ENGR3}
& &\tilde{\mathbf T}_{\mu }{}^{\nu }\stackrel{\mbox{\scriptsize def}}{=}
\delta_{\mu }{}^{\nu }{\mathbf L}-{\mathbf F}_{k}{}^{\lambda \nu }
A^{k}{}_{\lambda ,\mu }-\frac{\partial {\mathbf L}}
{\partial \phi^{A}{}_{,\nu }}\phi^{A}{}_{,\mu }
\nonumber \\
& &\mbox{\hspace*{7ex}}-
{\mathbf F}_{k}{}^{\nu }\psi^{k}{}_{,\mu }\; ,\\
\label{eq:def-ENGR4}
& &{\mathbf \Psi}_{\lambda}{}^{\mu \nu }
\stackrel{\mbox{\scriptsize def}}{=}
{\mathbf F}_{k}{}^{\mu \nu }A^{k}{}_{\lambda}=
-{\mathbf \Psi}_{\lambda}{}^{\nu \mu }\; ,
\end{eqnarray}
respectively. 
The conservation laws (\ref{eq:diff-consv-3}) and 
(\ref{eq:diff-consv-4}) remain unchanged.

In Ref.\cite{Kawai-Toma}, we have examined 
$M_{k},S_{kl},M^{c}{}_{\mu }$, and $L_{\mu }{}^{\nu }$ defined in the 
same ways as in \={P}.G.T.
for asymptotically flat space-time by choosing $\psi^{k}$ as given by 
Eq. (\ref{eq:psi-asymp}) and assuming some additional conditions on 
asymptotic behaviors of field variables. The same expressions as 
Eqs. (\ref{eq:em}) $\sim$ (\ref{eq:ext-orb-ang}) hold also in E.N.G.R.
\subsection{Transformation properties of energy-momentum and angular 
momentum densities}
\label{subsubsec-trans-ENGR}
In the case of E.N.G.R., the energy-momentum and \lq \lq spin" angular 
momentum densities ${}^{G}{\mathbf T}_{k}{}^{\mu }, 
{}^{M}{\mathbf T}_{k}{}^{\mu }, {}^{G}{\mathbf S}_{kl}{}^{\mu }$ and 
${}^{M}{\mathbf S}_{kl}{}^{\mu }$ in \=P.G.T. reduce to
\cite{Andrade-Guillen-Pereira} 
\begin{eqnarray}
\label{eq:em-G-ENGR}
{}^{G}{\mathbf T}_{k}{}^{\mu }
&=&\frac{\partial {\mathbf L}^{T}}{\partial\psi^{k}{}_{,\mu }}
=\frac{\partial{\mathbf L}^{T}}{\partial A^{k}{}_{\mu }}\; ,\\
\label{eq:em-M-ENGR} 
 {}^{M}{\mathbf T}_{k}{}^{\mu }&=&
\frac{\partial {\mathbf L}^{M}}{\partial \psi^{k}{}_{,\mu }}
+i\frac{\partial{\mathbf L}^{M}}
{\partial \phi^{A}{}_{,\mu }}(P_{k}\phi)^{A}
=\frac{\partial{\mathbf L}^{M}}{\partial A^{k}{}_{\mu }}\; ,\\ 
\label{eq:spin-G-ENGR}
{}^{G}{\mathbf S}_{kl}{}^{\mu }
&=&
-2\frac{\partial {\mathbf L}^{T}}
{\partial \psi^{[k}{}_{,\mu }}\psi_{l]}
-2{\mathbf F}_{[k}{}^{\nu \mu }A_{l]\nu }\; , \\ 
\label{eq:spin-M-ENGR}
{}^{M}{\mathbf S}_{kl}{}^{\mu }
&=&-2\frac{\partial {\mathbf L}^{M}}
{\partial \psi^{[k}{}_{,\mu }}\psi_{l]}
-i\frac{\partial {\mathbf L}^{M}}
{\partial \phi^{A}{}_{,\mu }}(M_{kl}\phi)^{A}\; .
\end{eqnarray}
Under the product transformations of Eqs. (\ref{eq:Lorentz-gauge-inf}) 
and (\ref{eq:coord-trans-inf}), these densities transform 
according to 
\widetext
\begin{eqnarray}
\label{eq:em-G-trans-ENGR}
{}^{G}{\mathbf T}^{\prime }{}_{k}{}^{\mu }
&=&{}^{G}{\mathbf T}_{k}{}^{\mu }
-\omega_{k}{}^{l}{}^{G}{\mathbf T}_{l}{}^{\mu }
+\epsilon^{\mu}{}_{,\nu }{}^{G}{\mathbf T}_{k}{}^{\nu }
-\epsilon^{\lambda}{}_{,\lambda }{}^{G}{\mathbf T}_{k}{}^{\mu }
-\omega_{k}{}^{l}{}_{,\nu }{\mathbf F}_{l}{}^{\mu \nu }\; ,\\
\label{eq:em-M-trans-ENGR}
{}^{M}{\mathbf T}^{\prime }{}_{k}{}^{\mu }
&=&
{}^{M}{\mathbf T}_{k}{}^{\mu }
-\omega_{k}{}^{l}{}^{M}{\mathbf T}_{l}{}^{\mu }
+\epsilon^{\mu}{}_{,\nu }{}^{M}{\mathbf T}_{k}{}^{\nu }
-\epsilon^{\lambda}{}_{,\lambda }{}^{M}{\mathbf T}_{k}{}^{\mu }\; ,\\
\label{eq:spin-G-trans-ENGR}
{}^{G}{\mathbf S}^{\prime }{}_{kl}{}^{\mu }
&=&{}^{G}{\mathbf S}{}_{kl}{}^{\mu }
-\omega_{k}{}^{m}{}^{G}{\mathbf S}{}_{ml}{}^{\mu }
-\omega_{l}{}^{m}{}^{G}{\mathbf S}{}_{km}{}^{\mu }
-2t_{[k}{}^{G}{\mathbf T}{}_{l]}{}^{\mu }
+\epsilon^{\mu }{}_{,\nu }{}^{G}{\mathbf S}{}_{kl}{}^{\nu }
-\epsilon^{\lambda}{}_{,\lambda}
{}^{G}{\mathbf S}{}_{kl}{}^{\mu }
-2t_{[k,\nu }{\mathbf F}_{l]}{}^{\mu \nu }{}\; ,\nonumber \\
{}\\
\label{eq:spin-M-trans-ENGR}
{}^{M}{\mathbf S}^{\prime }{}_{kl}{}^{\mu }
&=&{}^{M}{\mathbf S}{}_{kl}{}^{\mu }
-\omega_{k}{}^{m}{}^{M}{\mathbf S}{}_{ml}{}^{\mu }
-\omega_{l}{}^{m}{}^{M}{\mathbf S}{}_{km}{}^{\mu }
-2t_{[k}{}^{M}{\mathbf T}{}_{l]}{}^{\mu }
+\epsilon^{\mu }{}_{,\nu }{}^{M}{\mathbf S}{}_{kl}{}^{\nu }
-\epsilon^{\lambda}{}_{,\lambda}
{}^{M}{\mathbf S}{}_{kl}{}^{\mu }\; .
\end{eqnarray}
\narrowtext
Equation (\ref{eq:canonical-grav}) reduces to
\begin{equation}
\label{eq:canonical-grav-ENGR}
{}^{G}\tilde{\mathbf T}_{\mu }{}^{\nu }
=\delta_{\mu }{}^{\nu }{\mathbf L}^{T}-
{\mathbf F}_{k}{}^{\lambda \nu }A^{k}{}_{\lambda ,\mu }
-\frac{\partial {\mathbf L}^{T}}{\partial \psi^{k}{}_{,\nu }}
\psi^{k}{}_{,\mu } 
\end{equation}
with ${\mathbf L}^{T}\stackrel{\mbox{\scriptsize def}}{=}
\sqrt{-g}L^{T}$, while  Eqs. (\ref{eq:canonical-matter}) $\sim $ 
(\ref{eq;tot=G+M}) remain unchanged.

The densities ${}^{G}\tilde{\mathbf T}_{\mu }{}^{\nu }$ and 
${}^{M}{\mathbf T}_{\mu }{}^{\nu }$ transform according to
\widetext
\begin{eqnarray}
\label{eq:can-em-G-trans-ENGR}
{}^{G}\tilde {\mathbf T}^{\prime }{}_{\mu }{}^{\nu }
&=&{}^{G}\tilde {\mathbf T}{}_{\mu }{}^{\nu }-
t^{k}{}_{,\lambda \mu }{\mathbf F}_{k}{}^{\lambda \nu }
+t^{k}{}_{,\mu }
\frac{\partial {\mathbf L}^{T}}{\partial \psi^{k}{}_{,\nu }}
+\omega^{kl}{}_{,\mu }\frac{\partial {\mathbf L}^{T}}
{\partial \psi^{[k}{}_{,\nu }}\psi_{l]}+
\epsilon^{\lambda}{}_{,\mu }{}^{G}\tilde{\mathbf T}_{\lambda }{}^{\nu }-
\epsilon^{\nu }{}_{,\lambda }{}^{G}\tilde{\mathbf T}_{\mu }{}^{\lambda }\nonumber \\
& &-\epsilon^{\lambda }{}_{,\lambda }{}^{G}
\tilde{\mathbf T}_{\mu }{}^{\nu }
+\epsilon^{\rho}{}_{,\lambda \mu }
{\mathbf \Psi}_{\rho}{}^{\lambda \nu}\; ,\\
\label{eq:can-em-M-trans-ENGR}
{}^{M}{\mathbf T}^{\prime }{}_{\mu }{}^{\nu }
&=&{}^{M}{\mathbf T}{}_{\mu }{}^{\nu }
+t^{k}{}_{,\mu }\frac{\partial {\mathbf L}^{M}}
{\partial \psi^{k}{}_{,\nu }}+\omega^{kl}{}_{,\mu }
\frac{\partial {\mathbf L}^{M}}
{\partial \psi^{[k}{}_{,\nu }}\psi_{l]}
+\epsilon^{\lambda}{}_{,\mu }
{}^{M}{\mathbf T}_{\lambda }{}^{\nu }
-\epsilon^{\nu }{}_{,\lambda }{}^{M}
{\mathbf T}_{\mu }{}^{\lambda }
-\epsilon^{\lambda }{}_{,\lambda }{}^{M}{\mathbf T}_{\mu }{}^{\nu }
\; , 
\end{eqnarray}
\narrowtext
\noindent under the product transformations of Eqs. (\ref{eq:Lorentz-gauge-inf})
and (\ref{eq:coord-trans-inf}).
\noindent The transformation properties of 
${}^{G}\tilde{\mathbf M}_{\lambda}{}^{\mu \nu }$ and of 
${}^{M}\tilde{\mathbf M}_{\lambda}{}^{\mu \nu }$ are given by 
the same forms as Eqs. (\ref{eq:ext-orb-G-trans}) and 
(\ref{eq:ext-orb-M-trans}), respectively, where, for the present 
case, ${}^{G}\tilde {\mathbf T}^{\prime }{}_{\mu }{}^{\nu }$ and 
${}^{M}{\mathbf T}^{\prime }{}_{\mu }{}^{\nu }$ are given by 
Eqs. (\ref{eq:can-em-G-trans-ENGR}) and (\ref{eq:can-em-M-trans-ENGR}), respectively. Equation (\ref{eq:Psi-trans}) is reduced to 
\begin{eqnarray}
\label{eq:Psi-trans-ENGR}
{\mathbf \Psi}{}^{\prime }{}_{\lambda }{}^{\mu \nu }
&=&{\mathbf \Psi}{}_{\lambda }{}^{\mu \nu }+
t^{k}{}_{,\lambda }{\mathbf F}_{k}{}^{\mu \nu }
-\epsilon^{\rho}{}_{,\lambda }{\mathbf \Psi}{}_{\rho }{}^{\mu \nu }
+\epsilon^{\mu}{}_{,\rho}{\mathbf \Psi}{}_{\lambda }{}^{\rho \nu }
\nonumber \\
& &+\epsilon^{\nu}{}_{,\rho}{\mathbf \Psi}{}_{\lambda }{}^{\mu \rho }
-\epsilon^{\rho}{}_{,\rho}{\mathbf \Psi}{}_{\lambda }{}^{\mu \nu }\; .
\end{eqnarray} 
{\section{SUMMARY AND DISCUSSIONS}
We have examined the transformations properties of energy-momentum 
densities and of angular momentum densities both in \=P.G.T. 
and in E.N.G.R.

Results can be summarized as follows:
\begin{description}
\item[{[1]}] Results in $\overline{\mbox{\rm Poincar\'{e}}}$ gauge 
theory (\=P.G.T.):
\begin{description}
\item[(1A)] From Eqs. (\ref{eq:em-G-trans-PGT})$\sim $
(\ref{eq:spin-M-trans-PGT}), we see that the densities 
${}^{G}{\mathbf T}_{k}{}^{\mu }, {}^{M}{\mathbf T}_{k}{}^{\mu },
{}^{G}{\mathbf S}_{kl}{}^{\mu }$ and ${}^{M}{\mathbf S}_{kl}{}^{\mu }$
are all space-time vector densities, i.e., they transform as vector 
densities under general coordinate transformations. Their 
transformation properties under internal 
$\overline{\mbox{\rm Poincar\'{e}}}$ transformations are summarized 
as follows:
\begin{description}
\item[(a)] The dynamical energy-momentum density 
${}^{G}{\mathbf T}_{k}{}^{\mu }$ of the gravitational field is 
invariant under {\em local} translations. It transforms as a vector 
under {\em global } $SL(2,C)$-transformations. But, it is not 
vectorial under {\em local} $SL(2,C)$-transformations.   
\item[(b)] The dynamical energy-momentum density 
${}^{M}{\mathbf T}_{k}{}^{\mu }$ of the matter field $\phi^{A}$ is 
invariant under {\em local} translations. It transforms as a vector 
under {\em local } $SL(2,C)$-transformations.
\item[(c)] Under {\em global} translations, the \lq \lq spin" angular momentum density ${}^{G}{\mathbf S}_{kl}{}^{\mu }$ of the 
gravitational field receives transformations which correspond to translations in internal space-time, and it transforms as a 
tensor under {\em global} $SL(2,C)$-transformations. 
But, it is not tensorial under {\em local } 
$\overline{\mbox{\rm Poincar\'{e}}}$ transformations. 
\item[(d)] The \lq \lq spin" angular momentum density 
${}^{M}{\mathbf S}_{kl}{}^{\mu }$ of the matter field $\phi^{A}$ is 
tensorial under {\em local } $\overline{\mbox{\rm Poincar\'{e}}}$ transformations. 
\end{description}
\item[(1B)] From Eqs. (\ref{eq:can-em-G-trans}) $\sim $ 
(\ref{eq:Psi-trans}), we see that 
${}^{G}\tilde{\mathbf T}_{\mu }{}^{\nu },$
${}^{M}{\mathbf T}_{\mu }{}^{\nu },
{}^{G}\tilde{\mathbf M}_{\lambda}{}^{\mu \nu }$, and 
${}^{M}\tilde{\mathbf M}_{\lambda}{}^{\mu \nu }$ are all invariant 
under {\em global} internal $\overline{\mbox{\rm Poincar\'{e}}}$ 
transformations. They are not invariant under {\em local}
$\overline{\mbox{\rm Poincar\'{e}}}$ transformations.
Also, we can see the following:
\begin{description}
\item[(e)] The canonical energy-momentum density 
${}^{G}\tilde{\mathbf T}_{\mu }{}^{\nu }$ of the gravitational field 
transforms as tensor densities under affine coordinate 
transformation $x^{\prime \mu }=a^{\mu }{}_{\nu }x^{\nu }+b^{\mu }$, 
but it does not transform as a tensor density under general 
coordinate transformations\cite{tilde}.
\item[(f)] The canonical energy-momentum density 
${}^{M}{\mathbf T}_{\mu }{}^{\nu }$ of the matter field $\phi^{A}$ 
transforms as a tensor density under general coordinate 
transformations.
\item[(g)] Both of \lq \lq extended orbital angular momentum" 
densities ${}^{G}\tilde{\mathbf M}_{\lambda}{}^{\mu \nu }$ and 
${}^{M}\tilde{\mathbf M}_{\lambda}{}^{\mu \nu}$ transform as tensor 
densities under {\em constant } $GL(4,R)$-coordinate transformations, 
and they receive space-time translations under {\em constant} 
coordinate transformations. They do not transform as tensor 
densities under general coordinate transformations.
\end{description}
\end{description}
\item[{[2]}] Results in extended new general relativity\\
 (E.N.G.R.):
\begin{description}
\item[(2A)] All the densities ${}^{G}{\mathbf T}_{k}{}^{\mu },
{}^{M}{\mathbf T}_{k}{}^{\mu },{}^{G}{\mathbf S}_{kl}{}^{\mu }$, 
and ${}^{M}{\mathbf S}_{kl}{}^{\mu }$ are space-time vector
densities. Also for the case of E.N.G.R., the same 
statements as ({\bf a}), ({\bf b}), and ({\bf d}) in ({\bf 1A}) 
hold true for 
${}^{G}{\mathbf T}_{k}{}^{\mu },{}^{M}{\mathbf T}_{k}{}^{\mu }$, 
and ${}^{M}{\mathbf S}_{kl}{}^{\mu }$. As for 
${}^{G}{\mathbf S}_{kl}{}^{\mu }$, we have the following
\cite{local-SL2C-non-tensorial}:
\begin{description}
\item[(c$^{'}$)] The density ${}^{G}{\mathbf S}_{kl}{}^{\mu }$ receives 
transformations which correspond to translations of the origin of 
internal space-time under {\em global} translations, and it 
transforms as a tensor under {\em local} $SL(2,C)$-transformations. 
But, it is not tensorial under {\em local} internal translations.
\end{description}
\item[(2B)] Also for ${}^{G}\tilde{\mathbf T}_{\mu }{}^{\nu },
{}^{M}{\mathbf T}_{\mu }{}^{\nu },
{}^{G}\tilde{\mathbf M}_{\lambda}{}^{\mu \nu }$, and 
${}^{M}\tilde{\mathbf M}_{\lambda}{}^{\mu \nu }$ in E.N.G.R., the 
same statements as in ({\bf 1B}) hold true. 
\end{description}
\end{description}
{\em Since ${}^{G}{\mathbf T}_{k}{}^{\mu },
{}^{M}{\mathbf T}_{k}{}^{\mu }, {}^{G}{\mathbf S}_{kl}{}^{\mu }$,  
and ${}^{M}{\mathbf S}_{kl}{}^{\mu }$ are all space-time vector 
densities in both theories, the energy-momenta and angular momenta 
of the gravitational field and of the matter field are defined well, 
and  independent of the coordinate system employed.} For example, 
the energy-momentum ${}^{G}M_{k}$ of the gravitational field is 
defined by 
\begin{equation}
{}^{G}M_{k}\stackrel{\mbox{\scriptsize def}}{=}
\int_{\sigma}{}^{G}{\mathbf T}_{k}{}^{\mu }d\sigma_{\mu }\; ,
\end{equation}
and we have
\begin{equation}
{}^{G}M_{k}=
\int_{\sigma}{}^{G}{\mathbf T}_{k}{}^{\mu }d\sigma_{\mu }
=\int_{\sigma}{}^{G}{\mathbf T}^{\prime }{}_{k}{}^{\mu }
d\sigma^{\prime}{}_{\mu }\; .
\end{equation}

As we have mentioned in the final parts of Secs. \ref{subsec:outline-PGT} 
and \ref{subsec:reduction}, the total energy-momentum and 
the {\em total} (={\em spin}+{\em orbital}) angular momentum are 
given by $M_{k}$ and $S_{kl}$ for an asymptotically flat space-time,  
while the canonical energy-momentum $M^{c}{}_{\mu }$ and 
\lq \lq extended orbital angular momentum" $L_{\mu }{}^{\nu }$, 
which are obtained as the integrations of non-tensorial 
quantities $\tilde{\mathbf T}_{\mu }{}^{\nu }$ and 
$\tilde{\mathbf M}_{\lambda}{}^{\mu \nu }$, on the other hand, 
vanish and are trivial. 

In both in \=P.G.T. and in E.N.G.R., the 
densities ${}^{M}{\mathbf T}_{k}{}^{\mu }$ and 
${}^{M}{\mathbf S}_{kl}{}^{\mu }$ are well behaved under {\em local} 
internal $\overline{\mbox{\rm Poincar\'{e}}}$ transformations, while 
the energy-momentum density ${}^{G}{\mathbf T}_{k}{}^{\mu }$ of the 
gravitational field is well behaved under {\em local} internal 
translations. In E.N.G.R., the \lq \lq spin" angular momentum 
density ${}^{G}{\mathbf S}_{kl}{}^{\mu }$ of the gravitational field 
is well behaved under {\em local} internal $SL(2,C)$-transformations.

It is worth mentioning here that the Lagrangian densities 
$L^{G}+L^{M}$ in \=P.G.T. and $L^{T}+L^{M}$ in E.N.G.R. are both 
invariant under {\em local} internal translations and under 
{\em global} internal $SL(2,C)$-transformations, but they violate 
the invariance under {\em local} internal $SL(2,C)$-transformations.
 Thus, one may claim that we need not bother about the fact that 
${}^{G}{\mathbf T}_{k}{}^{\mu }$ in \=P.G.T. and in E.N.G.R. and 
${}^{G}{\mathbf S}_{kl}{}^{\mu }$ in \=P.G.T. are not tensorial 
under {\em local} internal $SL(2,C)$-transformations. In E.N.G.R., 
in particular, this can be strongly asserted, because 
{\em local} $SL(2,C)$-gauge invariance is rather accidental
\cite{probable-parameters} in this theory due to the lack of 
$SL(2,C)$-gauge potential $A^{kl}{}_{\mu }$.
  
We now give comments on alternative choices of sets of independent 
field variables:
\begin{description}
\item[{\{1\}}] In \=P.G.T., we can choose $\{\psi^{k},e^{k}{}_{\mu },
A^{kl}{}_{\mu },\phi^{A}\}$ as the set of independent field variables
\cite{Kawai-Saitoh01,Kawai-Saitoh02,PGT}. The dynamical and canonical 
energy-momentum densities and spin and \lq \lq extended orbital 
angular momentum" densities can be defined also for this case. The 
dynamical energy-momentum and spin angular momentum densities are 
space-time vector densities, and the canonical energy-momentum and 
\lq \lq extended orbital angular momentum" densities are not 
space-time tensor densities. The transformation properties of these 
quantities under the $\overline{\mbox{\rm Poincar\'{e}}}$ gauge 
transformations are much the same as in the case with 
$\{\psi^{k},A^{k}{}_{\mu },A^{kl}{}_{\mu },\phi^{A}\}$ being 
employed. For the choice $\{\psi^{k},e^{k}{}_{\mu }, 
A^{kl}{}_{\mu },\phi^{A}\}$, however, the total dynamical 
energy-momentum vanishes identically and the total canonical 
energy-momentum gives the total energy-momentum for the 
asymptotically flat space-time for a suitably chosen coordinate 
system. Also, for asymptotically flat space-time, the spin angular 
momentum and orbital angular momentum are both divergent, and the 
total angular momentum is obtainable only as the sum of spin and 
orbital angular momenta. Thus, {\em the total energy-momentum and 
the total angular momentum cannot be defined independently of the 
coordinate system employed}, because both of the canonical 
energy-momentum and orbital angular momentum densities are not 
tensor densities.
\item[{\{2\}}] In E.N.G.R., we can choose $\{\psi^{k},e^{k}{}_{\mu },
\phi^{A}\}$ as the set of independent field variables
\cite{Kawai-Toma,NGR}. For this choice, almost the same statements as 
in \{{\bf 1}\} hold, and the total energy-momentum and total 
angular momentum are obtained only by using densities which are not 
tensor densities.      
\end{description}
The choice $\{\psi^{k},A^{k}{}_{\mu },A^{kl}{}_{\mu },\phi^{A}\}$
with the condition (\ref{eq:psi-asymp}) in \=P.G.T. and the choice 
$\{\psi^{k},A^{k}{}_{\mu },\phi^{A}\}$ with the condition 
(\ref{eq:psi-asymp}) in E.N.G.R. are preferential to all the other 
choices.

In general relativity, all the known energy-momentum and angular 
momentum densities of the gravitational field are not space-time 
tensor densities. $\overline{\mbox{\rm Poincar\'{e}}}$ gauge theory 
and extended new general relativity are preferential to general 
relativity in the point that the former two theories have 
energy-momentum and angular momentum densities of the gravitational  
field which are true space-time vector densities\cite{GR}.
\appendix
\section{}
The irreducible components $t_{klm}$, $v_{k}$, and 
$a_{k}$ of $T_{klm}$ are defined by the following:
\begin{eqnarray}
\label{eq:irreducible-comp-torsion}
t_{klm}&\stackrel{\mbox{\scriptsize def}}{=}&
\frac{1}{2}(T_{klm}+T_{lkm})+\frac{1}{6}
(\eta_{mk}v_{l}+\eta_{ml}v_{k})\nonumber \\
& &-\frac{1}{3}\eta_{kl}v_{m}
\; ,\\
v_{k}&\stackrel{\mbox{\scriptsize def}}{=}&T^{l}{}_{lk}\; ,\\
a_{k}&\stackrel{\mbox{\scriptsize def}}{=}&
\frac{1}{6}\varepsilon_{klmn}T^{lmn}
\; ,
\end{eqnarray}
where the symbol $\varepsilon_{klmn}$ stands for completely 
anti-symmetric Lorentz tensor with $\varepsilon_{(0)(1)(2)(3)}=-1
$\cite{index-parentheses}.

The irreducible components $A_{klmn}, B_{klmn},C_{klmn}, E_{kl},$
$I_{kl}, R$ of $R_{klmn}$ are defined by the following: 
\begin{eqnarray}
\label{eq:irreducible-comp-curvature}
A_{klmn}&\stackrel{\mbox{\scriptsize def}}{=}&
\frac{1}{6}(R_{klmn}+R_{kmnl}+R_{knlm}+R_{lmkn}\nonumber \\
& &+R_{lnmk}+R_{mnkl})\; ,\\
B_{klmn}&\stackrel{\mbox{\scriptsize def}}{=}&
\frac{1}{4}(W_{klmn}+W_{mnkl}-W_{knlm}-W_{lmkn})\; ,\\
C_{klmn}&\stackrel{\mbox{\scriptsize def}}{=}&
\frac{1}{2}(W_{klmn}-W_{mnkl})\; ,\\
E_{kl}&\stackrel{\mbox{\scriptsize def}}{=}&
\frac{1}{2}(R_{kl}-R_{lk})\; ,\\
I_{kl}&\stackrel{\mbox{\scriptsize def}}{=}&
\frac{1}{2}(R_{kl}+R_{lk})-\frac{1}{4}\eta_{kl}R\; ,\\
R&\stackrel{\mbox{\scriptsize def}}{=}&\eta^{kl}R_{kl}
\end{eqnarray}
with 
\begin{eqnarray}
\label{eq:W-R}
W_{klmn}&\stackrel{\mbox{\scriptsize def}}{=}&
R_{klmn}-\frac{1}{2}(\eta_{km}R_{ln}
+\eta_{ln}R_{km}-\eta_{kn}R_{lm}\nonumber \\
& &-\eta_{lm}R_{kn})
+\frac{1}{6}(\eta_{km}\eta_{ln}-\eta_{lm}\eta_{kn})R\; ,\\
R_{kl}&\stackrel{\mbox{\scriptsize def}}{=}&\eta^{mn}R_{kmln}\; .
\end{eqnarray}


\begin{thebibliography}{1}
\bibitem{Poincare} The covering group $\bar{P}_{0}$ of the proper 
orthochronous Poincar\'{e} group will be called the 
$\overline{\mbox{\rm Poincar\'{e}}}$ group, and its elements will 
be called $\overline{\mbox{\rm Poincar\'{e}}}$ transformations.
\bibitem{index}Unless otherwise stated, we use the following 
conventions for indices: The middle part of the Greek alphabet, 
$\mu,\nu,\lambda,\dots$, refers to 0, 1, 2, and 3. In a similar way, 
the middle part of the Latin alphabet, $i,j,k,\dots$, means 0, 1, 2, 
and 3, unless otherwise stated. The capital letters $A$ and $B$ are 
used for indices for components of the field $\phi $, and $N$ 
denotes the dimension of the representation $\rho$.
\bibitem{derivative}For function $f$ on $M$, we define 
$f_{,\mu}\stackrel{\mbox{\scriptsize def}}{=}
\partial f/\partial x^{\mu }$.
\bibitem{symmetrization-antisymmetrization}We define 
\begin{eqnarray*}
A_{\cdots [\mu \cdots \nu]\cdots }
&\stackrel{\mbox{\scriptsize def}}{=}&\frac{1}{2}
(A_{\cdots \mu \cdots \nu \cdots }-
A_{\cdots \nu \cdots \mu \cdots })\; ,\\
A_{\cdots (\mu \cdots \nu)\cdots }
&\stackrel{\mbox{\scriptsize def}}{=}&\frac{1}{2}
(A_{\cdots \mu \cdots \nu \cdots }+A_{\cdots \nu \cdots \mu \cdots })\;.
\end{eqnarray*}
\bibitem{raising-lowering}The field components $e^{k}{}_{\mu }$
and $e^{\mu }{}_{k}$ are used to convert Latin and Greek indices,
similarly as in the case of $D_{k}\phi^{A}$ and $D_{\mu }\phi^{A}$.
Also, raising and lowering the indices $k,l,m,\dots $ are 
accomplished with the aid of $(\eta^{kl})
\stackrel{\mbox{\scriptsize def}}{=}(\eta_{kl})^{-1}$ and 
$(\eta_{kl})$.
\bibitem{Newtonian-limit}The requirement that the theory has the 
Newtonian limit restricts the parameter $d$ as $d=1/(2\kappa)$
\cite{Hayashi-Shirafuji01}. Here, $\kappa$ denotes the Einstein 
constant $\kappa =8\pi G/c^{4}$ with $G$ and $c$ being the Newton 
gravitational constant and the light velocity in the vacuum, 
respectively. If $3d_{2}+2d_{3}=0=d_{5}+12d_{6}$ and if the 
intrinsic spins of the source fields are negligible in addition, 
the Einstein equation is obtainable\cite{Hayashi-Shirafuji01}. 
In the discussions below, however, we do not impose any restriction 
on the parameters $c_{i},d_{j},d\;  
(i=1,2,3, j=1,2,3,\dots,6)$.
\bibitem{extended}Note that the anti-symmetric part $L_{[\mu \nu ]}$ 
is the generator of coordinate Lorentz transformations and is 
orbital angular momentum.
\bibitem{independent-set} In the rest of this section, the set 
$\{\psi^{k}, A^{k}{}_{\mu },\phi^{A}\}$ will be employed as the set 
of independent field variables. The case when $\{\psi^{k}, 
e^{k}{}_{\mu },\phi^{A}\}$ is employed will be mentioned in the 
final section.
\bibitem{tilde} The symbol $\;\tilde{}\;$ in the symbols 
${}^{G}\tilde{\mathbf T}_{\mu }{}^{\nu},{}^{G}
\tilde{\mathbf M}_{\lambda}{}^{\mu \nu }$, etc. is to express 
that these densities are not space-time tensor densities.
\bibitem{index-parentheses}Latin indices are put in parentheses to 
distinguish them from Greek indices.
\end{thebibliography}
\end{document}